%% file: cbass-backend-paper.tex
\documentclass[usenatbib]{mnras}
\usepackage{mathptmx}
\usepackage[T1]{fontenc}
\usepackage{graphicx}
\usepackage{multirow} 
\usepackage{color} 
\usepackage{amsmath,amssymb}
\usepackage{upgreek}

\newcommand{\EXP}[1]{\langle #1 \rangle}
\newcommand{\I}{\mathrm{i}}

\title[C-BASS digital backend]{The C-Band All-Sky Survey (C-BASS): Digital backend for the northern survey}

\author[M. A. Stevenson et al.]{%
M. A. Stevenson,$^{1}$ 
T. J. Pearson,$^1$\thanks{E-mail \url{tjp@astro.caltech.edu}} 
Michael~E.~Jones,$\!^{2}$
C.~J.~Copley,$\!^{2,3}$ 
C.~Dickinson,$\!^{4}$ 
\newauthor %
J.~J.~John,$\!^{2}$ 
O.~G.~King,$\!^{1,2}$ 
S.~J.~C.~Muchovej,$\!^{1}$
and Angela~C.~Taylor$^{2}$ 
 \\ 
$^{1}$California Institute of Technology, Pasadena, CA 91125, USA \\
$^{2}$Sub-department of Astrophysics, University of Oxford, Denys Wilkinson Building, Keble Road, Oxford OX1 3RH, UK \\
$^{3}$Department of Physics and Electronics, Rhodes University, Drostdy Road, Grahamstown, 6139, South Africa \\
$^{4}$Jodrell Bank Centre for Astrophysics, School of Physics and Astronomy, The University of Manchester, Manchester, M13 9PL, UK \\
}

\date{Accepted XXX. Received YYY; in original form ZZZ}

\pubyear{2018}

\begin{document}
\label{firstpage}
\pagerange{\pageref{firstpage}--\pageref{lastpage}}
\maketitle

\begin{abstract}
  The C-Band All-Sky Survey (C-BASS) is an all-sky full-polarization
  survey at a frequency of 5\,GHz, designed to provide data complementary
  to the all-sky surveys of \textit{WMAP} and \textit{Planck}
  and future CMB $B$-mode polarization imaging surveys. We describe the
  design and performance of the digital backend used for the northern
  part of the survey. In particular we describe the features that efficiently implement the demodulation and filtering required to suppress contaminating signals in the time-ordered data, and the capability for real-time correction of detector non-linearity and receiver balance. 
\end{abstract}




\section{Introduction}

The C-Band All-Sky Survey (C-BASS) is an all-sky full-polarization survey at a frequency of 5\,GHz, designed to provide data complementary to the all-sky surveys of \textit{WMAP} and \textit{Planck} and future cosmic microwave background imaging surveys \citep{projectpaper}. It uses two single-dish radio telescopes, one in the northern hemisphere at the Owens Valley Radio Observatory, California, and one in the southern hemisphere at the South African Radio Astronomy Observatory site at Klerefontein. Although the telescopes have different optics they have matched beams with FWHM 45\,arcmin. Both telescopes are equipped with similar  dual-polarization receivers that combine a continuous-comparison radiometer for measuring the Stokes parameter combinations $I+V$ and $I-V$ with a correlation polarimeter for measuring Stokes $Q$ and $U$ by cross-correlation of the two circularly polarized signals. However, the two receivers differ in the implementation of the radiometer and polarimeter. 

The northern receiver \citep{2014MNRAS.438.2426K} uses analogue electronics to process the 4.5 -- 5.5 GHz radio frequency (RF) signals in a single frequency channel, forming linear combinations of the orthogonal polarization voltages and reference signals using passive microwave circuits, and ultimately producing analogue outputs whose powers are detected with Schottky diodes. The newer southern receiver (C. Copley et al., in preparation), in contrast, samples the voltages using high-speed digitizers, and all the subsequent signal combination, power detection and integration are done using digital signal processing.

In this paper we present the design of the digital backend for the northern receiver, which was operated between 2009 and 2015. This system provided the sampling, data acquisition and read-out, as well as the phase-switching modulation scheme that was essential for the suppression of systematic effects in the receiver. Section~\ref{sec:context} summarizes the design of the northern receiver and the requirements it places on the backend.
Section~\ref{backend_hardware.sec} describes the backend hardware, and Section~\ref{sec:backend_firmware} describes the custom field-programmable gate array (FPGA) firmware developed to carry out the demodulation and integration and pass the data to a data-acquisition computer. We emphasize the novel filtering and timing algorithms needed to carry out the processing in the FPGA environment. Section~\ref{sec:conclusions} presents the conclusions.

\section{Context and Requirements}
\label{sec:context}

\subsection{Receiver architecture}

\begin{figure*}
\includegraphics[width=\textwidth]{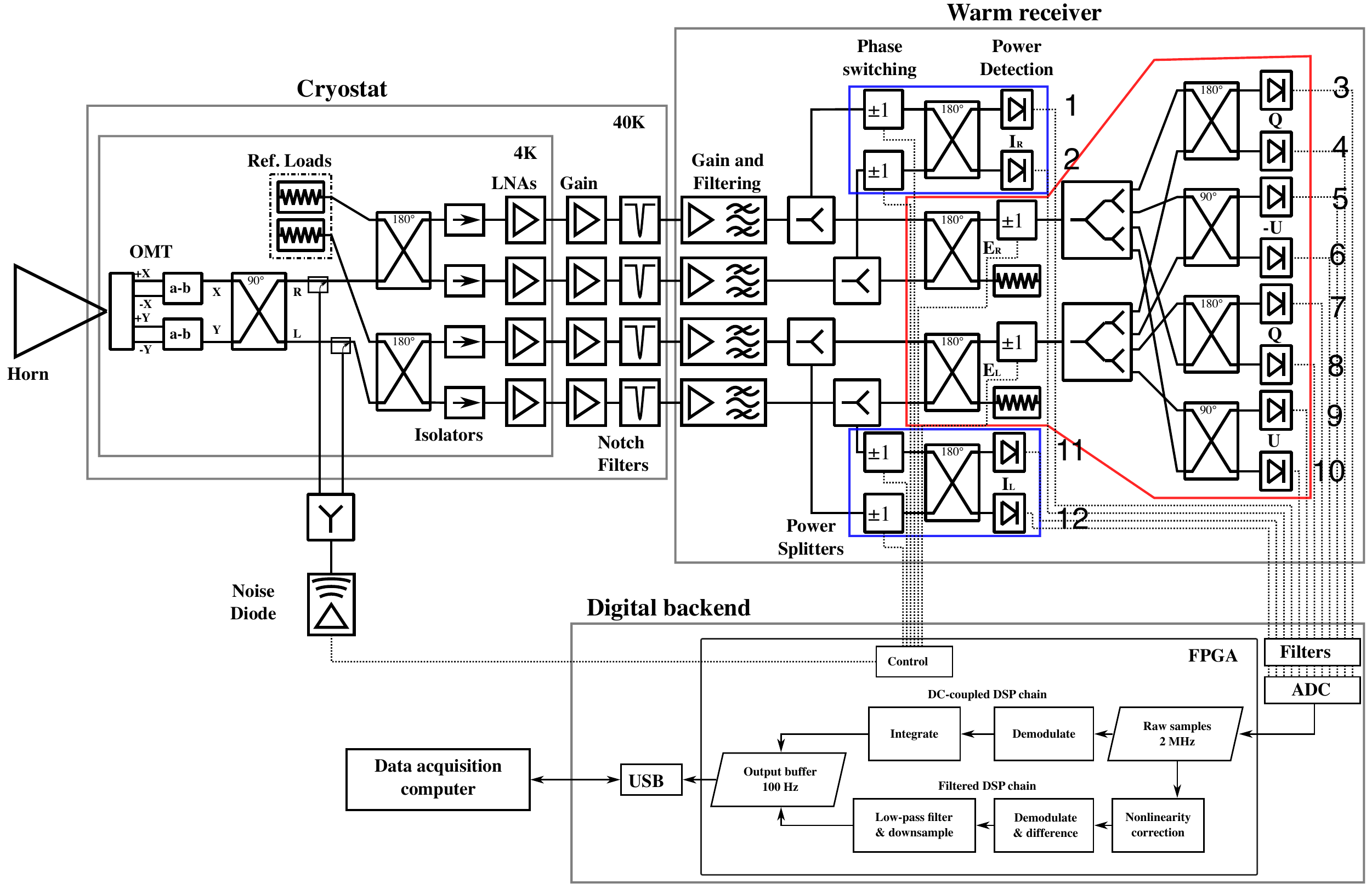}
\caption{C-BASS north receiver system diagram. The backend described in this paper is represented by the box at the bottom containing the FPGA. Adapted from \citet{2014MNRAS.438.2426K}.\label{system-diag}}
\end{figure*}

The C-BASS North receiver is a continuous-comparison radiometer combined with a correlation polarimeter \citep{projectpaper,2014MNRAS.438.2426K}. The radiometer architecture is similar to those of {\it WMAP} \citep{2003ApJS..145..413J} and the {\it Planck} Low-Frequency Instrument \citep{2002A&A...391.1185S}. The receiver consists of a cryogenic section containing the cold radio-frequency (RF) components, a warm receiver section of analogue RF components which implements the radiometer and polarimeter, and a digital backend which is the subject of this paper. The system estimates the Stokes parameters $I,Q,U,V$ from the complex voltages representing right and left circular polarization $E_L$ and $E_R$ using \citep[e.g.,][]{1996A&AS..117..161H,2014JKAS...47...15T}
\begin{align}
I &= \EXP{E_R E_R^*} + \EXP{E_L E_L^*} \\
Q &=  \EXP{E_R E_L^*} + \EXP{E_L E_R^*}\\ 
U &=  -\I[\EXP{E_R E_L^*} - \EXP{E_L E_R^*}]\\
V &= \EXP{E_R E_R^*} - \EXP{E_L E_L^*}\,.
\end{align}
Total intensity $I$ is measured relative to a pair of internal reference loads.  Fig \ref{system-diag} shows the overall system diagram.   

The sky signal enters the system through a corrugated feed horn, and two orthogonal linear polarizations are extracted and converted to circular polarization $E_L$ and $E_R$. A noise diode calibration signal is injected with equal amplitudes in to each of these signal paths. Two comparison signals from cold loads, $E_{\rm refR}$ and $E_{\rm refL}$ are combined with the sky signals using two $180\degr$ hybrids to create the four combinations $E_R \pm E_{\rm refR}$ and $E_L\pm E_{\rm refL}$, which are amplified with cooled low-noise amplifiers (LNAs) and filtered to remove out-of-band and in-band interfering signals. 

The warm receiver combines these four RF signals to form twelve data ``channels.'' The power in each channel is measured with a Schottky square-law detector diode and sampled and digitized in the backend. The twelve channels are labelled $S_1$ to $S_{12}$ in Fig.~\ref{system-diag}; $S_1$, $S_2$, $S_{11}$ and $S_{12}$ are the radiometer outputs and $S_3$--$S_{10}$ are the polarimeter outputs. In the radiometer part  (\textit{blue boxes} in Fig. \ref{system-diag}), the four RF signals are passed through $180\degr$ phase switches which modulate the signals at 1~kHz. A $180\degr$ hybrid then separates the signals into $E_R$, $E_{\rm refR}$, $E_L$, and $E_{\rm refL}$, and the square-law detectors generate voltages that are nominally proportional to
\begin{align}
S_1^+ &= \EXP{E_R E_R^*},&\quad& S_1^- = \EXP{E_{\rm refR}E_{\rm refR}^*}\\
S_2^+ &= \EXP{E_{\rm refR}E_{\rm refR}^*},&\quad& S_2^- = \EXP{E_R E_R^*} \\
S_{11}^+ &= \EXP{E_L E_L^*},&\quad & S_{11}^- = \EXP{E_{\rm refL}E_{\rm refL}^*}\\
S_{12}^+ &= \EXP{E_{\rm refL}E_{\rm refL}^*},&\quad& S_{12}^- = \EXP{E_L E_L^*} 
\end{align}
where the $+$ and $-$ superscripts refer to the two states of the phase-switch cycle. These signals therefore encode the powers in the right and left circular polarizations, and in the reference signals, measured through the same analogue signal chains.

In the polarimeter part (\textit{red box}), the reference load signals are discarded and the $E_L$ and $E_R$ voltages are combined in $180\degr$ and $90\degr$ hybrids to form $E_L \pm E_R$ and $E_L \pm \I E_R$, so the detector voltages are nominally
\begin{align}
S_3^\pm &= \EXP{|E_L \pm E_R|^2} = I \pm Q\\
S_4^\pm &= \EXP{|E_L \mp E_R|^2} = I \mp Q\\
S_5^\pm &= \EXP{|E_L \pm \I E_R|^2} = I \mp U \\
S_6^\pm &= \EXP{|E_L \mp \I E_R|^2} = I \pm U
\end{align}
These signals comprise the correlation between right and left circular polarization, modulated by the phase switch, and terms proportional to total powers, which are not modulated. Channels 7--10 nominally duplicate channels 3--6 and are not independent, but the data pass through different hybrids and detectors allowing for checks on some potential systematic errors.

By taking pairwise differences of the detector voltages, estimates of the linear-polarization Stokes parameters can now be formed as follows:
\begin{align}
S_1 - S_2 &= I_1 \equiv (I+V)-E^2_{\rm refR} \label{eq:diff1}\\
(S_3 - S_4)^\pm &= \pm Q_1 \\
(S_5 - S_6)^\pm &= \mp U_1 \\
(S_7 - S_8)^\pm &= \pm Q_2 \\
(S_9 - S_{10})^\pm &= \pm U_2 \\
S_{11} - S_{12} &= I_2 \equiv (I-V)-E^2_{\rm refL} \label{eq:diff6}
\end{align}
This is the primary function of the digital backend. Departures of the receiver from the ideal system summarized here are discussed in detail by \citet{2015MNRAS.446.1252K}. In summary, the effect of the phase switching and differencing is to remove to first order the effects of gain and phase imbalance in the analogue signal paths.

\subsection{Digital backend architecture}
\label{sec:backend-architecture}

The voltages representing the detected powers need to be sampled, demodulated, filtered, and integrated. This is done using a backend system based on an FPGA. The intrinsic bandwidth of the astronomical signal is less than $40\,\mathrm{Hz}$, set by the time required for the telescope to scan across its beam size. The phase switch modulation needs to be at significantly higher frequency than this, in order that the astronomical signal is effectively constant over a single phase switch cycle, and in addition the phase switch frequency has to be consistent with the data rates and decimation factors in the digital filtering (see Sect. \ref{backend_filtering.sec}). We chose a modulation frequency of 1 kHz, which modulates the astronomical data up to a band centred on 1 kHz and its odd multiples. The post-detection signal bandwidth, and hence the digitization sampling rate, are set such that the 1 kHz square-wave modulation is transmitted at $\sim 0.1$ per cent accuracy. Each of the twelve channels $S_1$--$S_{12}$ is anti-alias filtered with a 3\,dB bandwidth of $0.85\,\mathrm{MHz}$, allowing a sampling frequency of $2\,\mathrm{MHz}$ with minimal aliasing.

Following digitization, the signals are corrected for non-linearity of the detector diodes using look-up tables determined from laboratory measurements of the individual diodes. The signals are then differenced to produce the required Stokes signals (equations \ref{eq:diff1}--\ref{eq:diff6}), and the 1\,kHz phase switch modulation is demodulated. This reduces the science signal to baseband while mixing contaminating signals from, e.g., the 60-Hz power-supply voltage and its harmonics to higher frequencies. A chain of decimating low-pass filters follows, reducing the data rate to 100\,Hz and applying a 50\,Hz rectangular low-pass filter to the data. All signals away from this narrow band around the modulation frequencies are heavily suppressed by the filtering. However, this also means that the information on the DC levels of the individual detectors is lost. This information is useful to assess the state of balance of the receiver, i.e., the absolute power levels of the intensity reference and the sky brightness, and the relative gains of the branches of the continuous-comparison polarimeter. We therefore also store the integrated values of the twelve detector diodes in each of their phase-switch states over each 10 ms sample -- an additional 24 numbers per sample. These data are referred to as `unfiltered' data, in contrast to the six `filtered' data values.

Data are transferred to the control computer using a USB link limited to 25 transfers of 512\,B per second, or a transfer rate of  $12.5\,\mathrm{kB/s}$. This is however comfortably enough bandwidth to handle the integrated data. Given the sample rate of 100\,Hz (10\,ms samples), we therefore have 125\,B available for each sample to transfer the six science data values ($I_1, I_2, Q_1, Q_2, U_1, U_2$), 24 unfiltered samples, plus diagnostics. 

The functioning of the receiver can be monitored through dia\-gnostic signals, a number of which are available to the backend.  These include the status of the GPS time signal and the state of the internal memory.  Packaging these signals with the data time series facilitates monitoring the backend. As an additional diagnostic mode, it is possible to record short stretches of raw ADC at the full sample rate of 2 MHz (`burst mode'), and send these to the host computer.

\section{Hardware}\label{backend_hardware.sec}

The hardware used was based on an ADC/FPGA board designed at the University of Oxford for the LiCAS experiment \citep{Reichold:2006p154}. New DSP firmware in the FPGA was written for C-BASS, along with modifications to the LiCAS USB interfacing software.  

The backend hardware consists of four cards housed in a single 6U-height, 19-inch rack assembly.  An attenuator board allows for incoming signals to be reduced in amplitude to avoid ADC saturation if necessary. An analogue filter board provides the anti-aliasing filters and voltage gain to match the signal amplitude to the dynamic range of the ADCs. This board also converts the signals from single-ended to differential, in order to drive the differential-input ADCs. The analogue signals are conducted to the digital processing card via a backplane, where they are digitized by the ADCs and sent to the FPGA for processing.  The filtered and integrated signals are retrieved directly from the digital card via universal serial bus (USB). An interface card provides direct access to several FPGA input and output (I/O) pins, allowing for interfacing to control and timing signals.

\begin{figure*}
\begin{center}
\includegraphics[width=\textwidth]{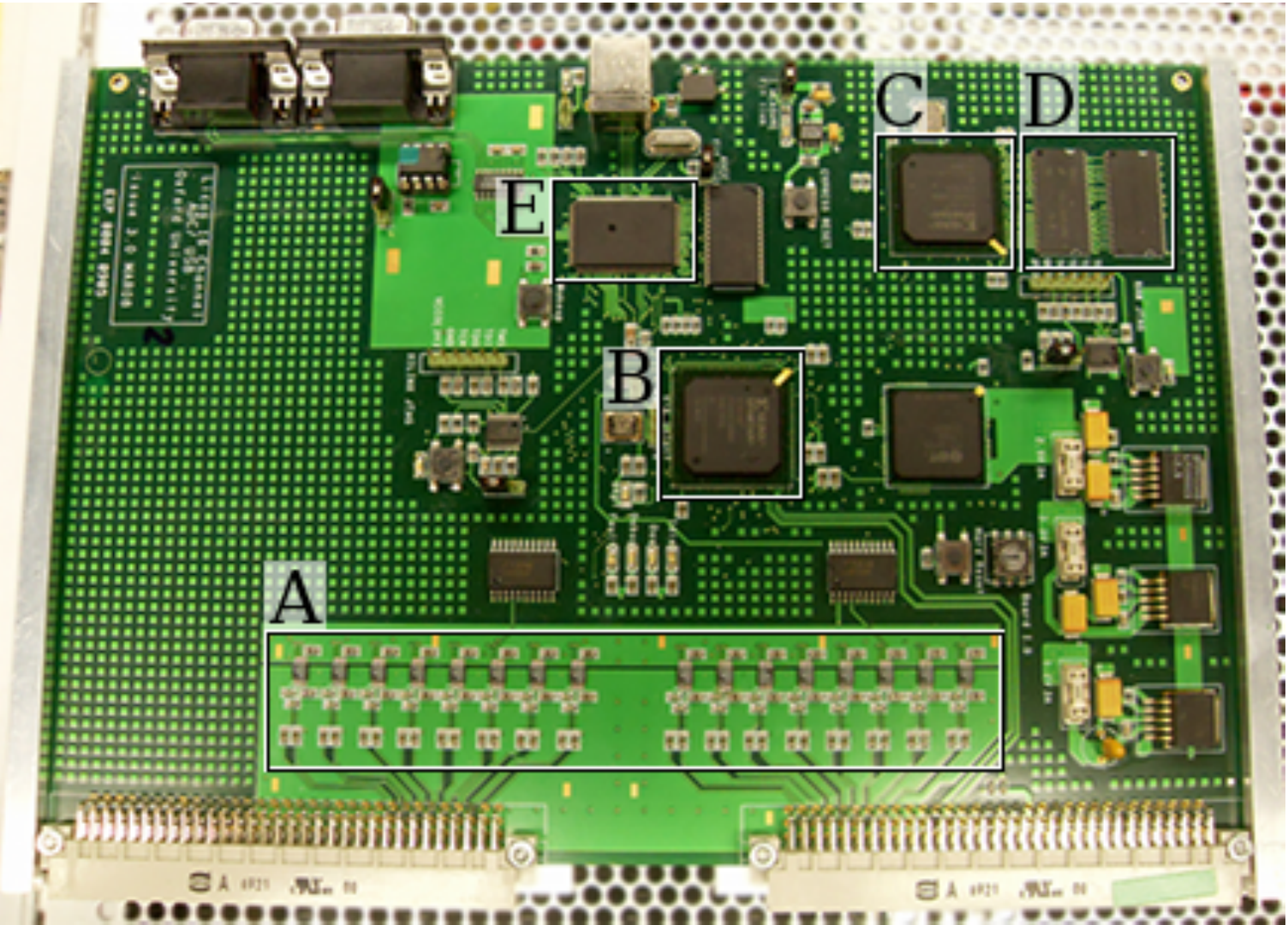}
\caption{The C-BASS digital processing (ADC/FPGA) card.  Labelled components are (A) the 16 ADCs, (B) the main FPGA, (C) the secondary FPGA, (D) the SDRAM chips, and (E) the USB microcontroller.}
\label{backend_digitalcard.fig}
\end{center}
\end{figure*}

The digital processing card is shown in Fig.~\ref{backend_digitalcard.fig}. This card includes (A) the 16 parallel ADCs, 12 of which are used, (B) the main FPGA, (C) the secondary FPGA, (D) $128\,\mathrm{MB}$ of synchronous dynamic random-access memory (SDRAM), and (E) the USB microcontroller.  The card is clocked at $50\,\mathrm{MHz}$, which is generated by an on-board oscillator.   The filtering and integration occur on the main FPGA. The secondary FPGA acts as an interface to the SDRAM, itself only used for the occasional collection of raw ADC outputs in burst mode.  Table~\ref{backend_chips.tab} lists the important chips and part numbers.

\input{Tables/backend_chips.tex}

The ADCs provide $14$-bit  differential measurements of signals in the range $\pm1.25\,\mathrm{V}$, with a maximum sampling rate of $2.78\,\mathrm{MS/s}$ (million samples per second)  when clocked at $50\,\mathrm{MHz}$.  They use an external control signal to initiate each conversion, allowing fine control of the exact sampling rate.  Conversion and serial transfer of a single reading requires 18 clock cycles (setting the maximum sampling rate).  For C-BASS, samples are requested every 25 clock cycles, giving a sampling rate of $2\,\mathrm{MS/s}$.

The main FPGA is a Xilinx Spartan 3,\footnote{Xilinx Inc.\ \url{https://www.xilinx.com}} which interfaces to and controls the ADCs, can communicate with the secondary FPGA when desired, and sends the processed data products to the USB microcontroller.  It has relatively few resources for an FPGA: twenty-four $18\rm bit\times 18\rm bit$ multipliers, twenty-four $18\,\mathrm{kb}$ blocks of RAM (which share inputs and outputs with their adjacent multipliers), and $17,280$ equivalent logic cells.  However this is sufficient to implement the signal processing operations we require.  A description of the necessary firmware is provided in Section~\ref{sec:backend_firmware}.

The SDRAM consists of two identical $64\,\mathrm{MB}$ chips, and requires buffering during reading and writing, provided by the secondary FPGA.  This FPGA is identical to the main FPGA in terms of resources, but with firmware tuned to this more modest role.  The $128\,\mathrm{MB}$ of memory is not used in normal operation, but is used to buffer raw ADC outputs when needed.

Transfer of processed data to the control computer is achieved via USB.  Twenty-five 512-byte packets are sent per second via USB 2.0 bulk transfers.  These packets are assembled on the main FPGA and sent to the USB microcontroller in 16-bit segments.  In the USB microcontroller, they are reassembled and made available upon request to the control computer.  The number of transfers per second is limited by the control computer, which is multitasking several parallel control processes and cannot be relied upon to request more than 25 transfers per second.

A number of I/O signals interface with the main FPGA through the digital I/O board.  These are the 1 pulse-per-second (1\,PPS) signal provided by a global positioning system (GPS) receiver, used for synchronizing the integrated data products with the antenna pointing and diagnostic data that are recorded separately by the control system; the six $180\degr$ phase switches in the analogue receiver; and the on/off switch for the calibration noise diode.

\section{Firmware}
\label{sec:backend_firmware}

\begin{figure*}
\begin{center}
\includegraphics[width=\textwidth]{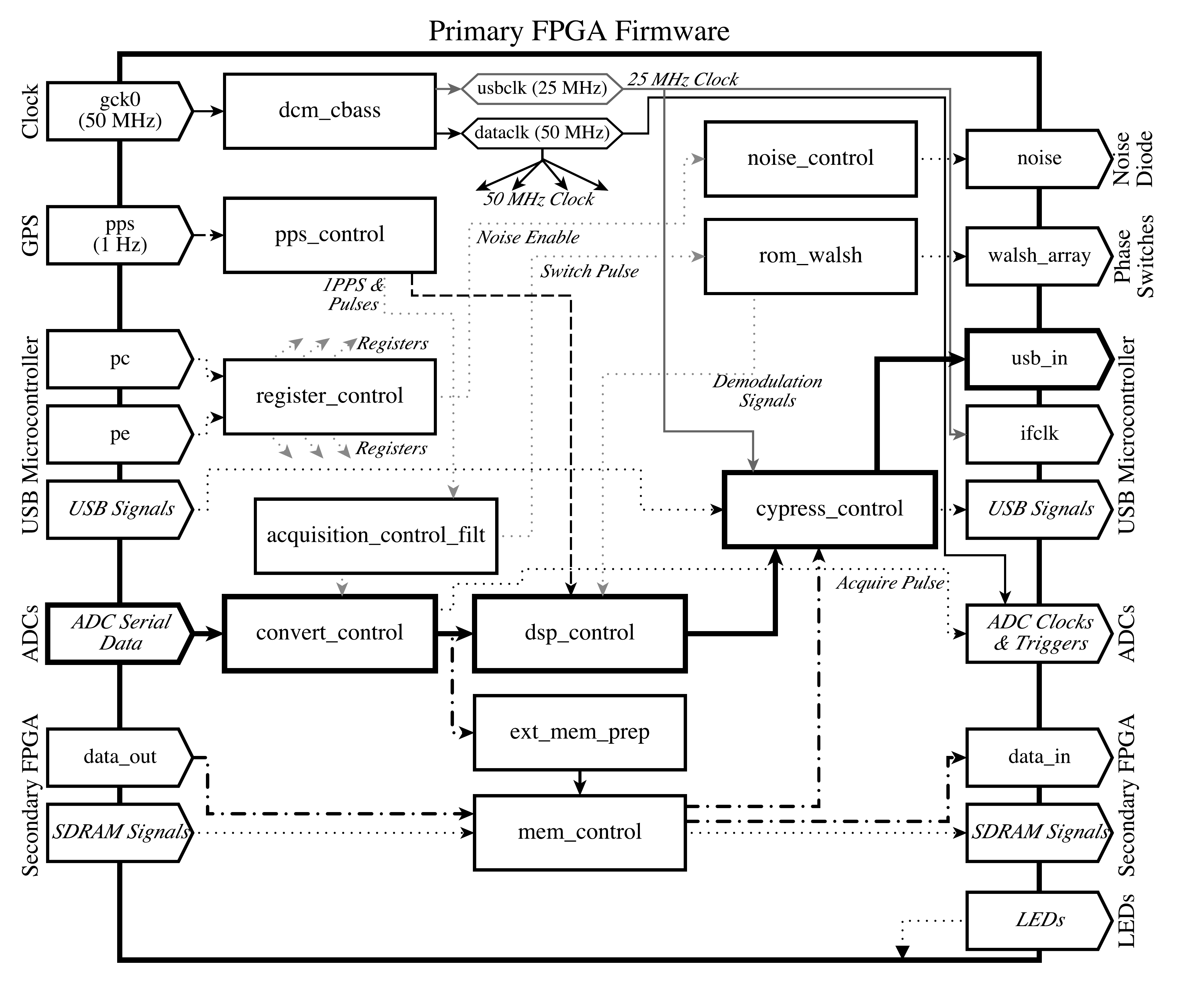}
\caption{Schematic diagram of the primary FPGA firmware.  The components are represented as rectangles.  Pentagons represent the input and output signals.  Internal signals, and groups thereof, are represented by arrows.  The bold lines and components mark the path of continuous data.  The dot-dashed line is for burst (high time-resolution) data.}
\label{backend_firmwareschematic.fig}
\end{center}
\end{figure*}

A schematic diagram of the FPGA firmware, including its most important components, is shown in Fig.~\ref{backend_firmwareschematic.fig}. All calculations in the FPGA are handled in two's complement fixed-point arithmetic.
The FPGA was programmed using VHDL (Very High Speed Integrated Circuit Hardware Description Language). The chief firmware components are:
\begin{description}
\item[\texttt{dcm\_cbass}] Provides the clocking signals for the entire FPGA, as well as for the ADCs and the USB microcontroller (see  Section~\ref{backend_timing.sec}).
\item[\texttt{pps\_control}] Contains algorithms for detecting the GPS 1PPS rising edge in the presence of noise.  Provides GPS-synchronized timing pulses to the data processing components (see Section~\ref{backend_pps.sec}).
\item[\texttt{register\_control}] Manages the variety of firmware registers used by the FPGA.
\item[\texttt{noise\_control}] Controls the noise diode on/off signal  (see Section~\ref{backend_firmware_control.sec}).
\item[\texttt{acquisition\_control\_filt}] Provides the timing pulses for ADC triggering, data trimming, and phase switching.
\item[\texttt{rom\_walsh}] Accepts timing pulses from \texttt{acquisition\_control\_filt} to generate the phase switching and demodulation signals (see Section~\ref{backend_firmware_control.sec}).
\item[\texttt{convert\_control}] Sends acquisition pulses to the ADCs and receives the serial data therefrom.  Provides the parallel, time series data to the various DSP options (see Section~\ref{backend_dsp.sec}).
\item[\texttt{dsp\_control}] Filters and integrates the time series data (see Section~\ref{backend_dsp.sec}).
\item[\texttt{cypress\_control}] Sends the processed data to the USB microcontroller.
\item[\texttt{ext\_mem\_prep}] When a burst of raw ADC data is requested, prepares the data for temporary storage in SDRAM.
\item[\texttt{mem\_control}] Sends data to the SDRAM when a burst of data is requested.  After the data have been collected, forwards the data from the SDRAM to \texttt{cypress\_control} to be sent out via USB.
\end{description}

\subsection{Timing}\label{backend_timing.sec}

The main FPGA is responsible for coordinating several different time bases:  the backend clock, the USB microcontroller clock, the ADC sampling rate, the phase switch rate, the integration timescale, the USB transfer rate, and the 1\,PPS from the GPS.  Although there are several constraints on these time bases, only the backend clock, the USB clock,  and 1PPS timescales are absolutely fixed.  The time bases were thus chosen to satisfy a number of constraints:
\begin{itemize}
\item the ADC sampling rate cannot exceed $2.78\,\mathrm{MS/s}$, 
\item the USB transfer rate cannot exceed $25$ bulk transfers per second, 
\item there must be an integral number of phase switch periods per integration, 
\item the phase switch frequency must be much less than the analog video bandwidth, and 
\item the higher frequencies need to be multiples of the lower.
\end{itemize}
The chosen frequencies and periods for the various timescales are summarized in Table~\ref{backend_timescales.tab}.

\input{Tables/backend_timescales.tex}

Although the FPGA can phase-lock its internal $25\,\mathrm{MHz}$ and $50\,\mathrm{MHz}$ clocks to the on-board $50\,\mathrm{MHz}$ clock (using its Digital Clock Manager, or DCM),  there is no way to phase-lock these clocks to the 1\,PPS.  Any variation in the backend clock thus goes uncorrected. 
Careful testing showed that the backend clock frequency is stable to $1\,\mathrm{ppm}$, but with a typical absolute frequency error of 5~ppm (250 Hz in 50 MHz).  The interface between these backend and GPS clock domains occurs at the integration stage, where the integration duration becomes locked to the 1\,PPS time base. This means that although the number of ADC samples can differ slightly between integrations there is negligible uncertainty in the effective integration time. 

The FPGA clock is divided into two further clock domains, the $50\,\mathrm{MHz}$ main clock and the $25\,\mathrm{MHz}$ USB microcontroller clock.  The two are bridged through one of the FPGA's RAM blocks, with the necessary configuration provided via a Xilinx-provided drop-in component.  This is discussed further in Section~\ref{backend_dsp.sec}.

\subsection{1\,PPS Handling}\label{backend_pps.sec}

\begin{figure*}
\begin{center}
\includegraphics[width=\textwidth]{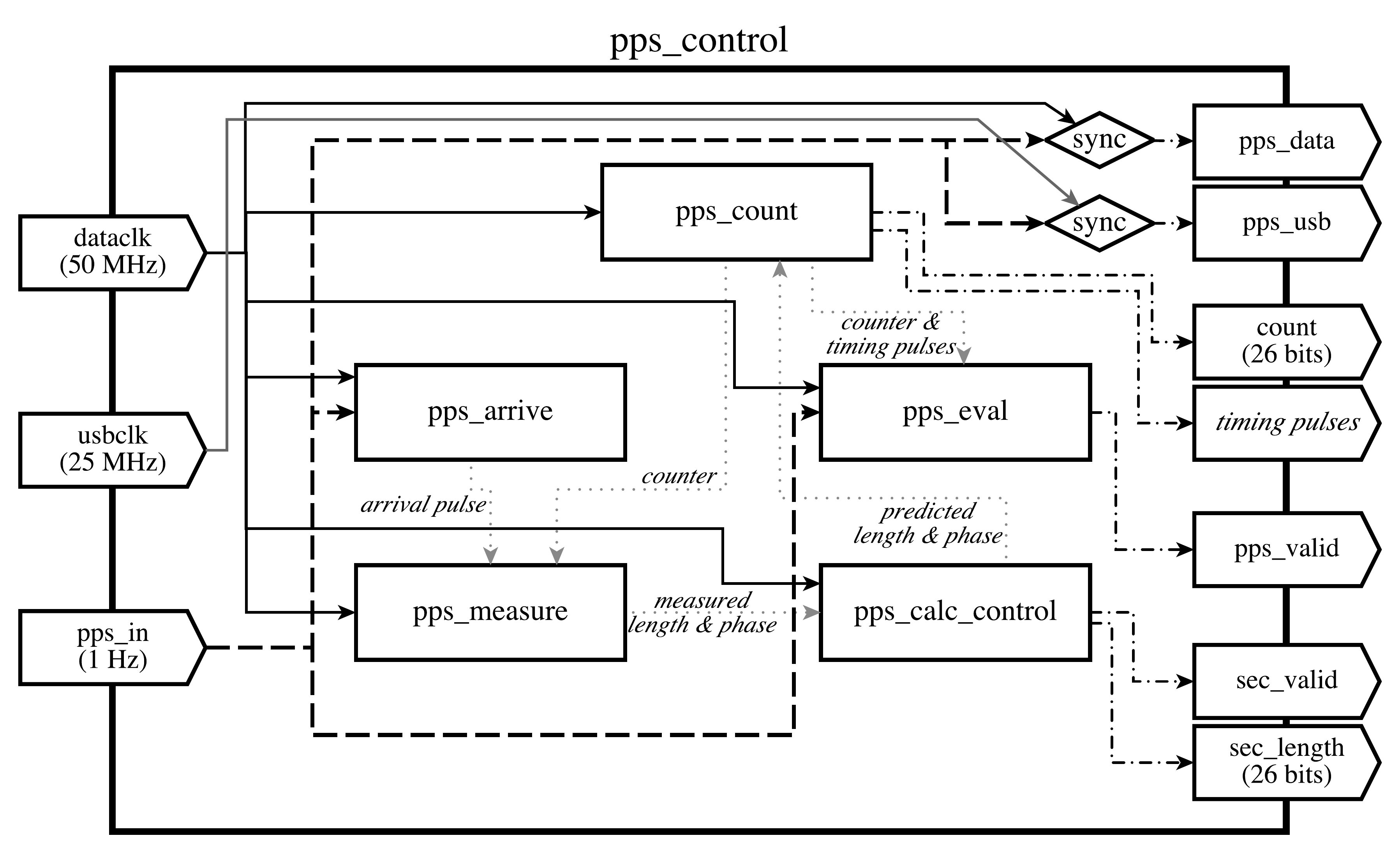}
\caption{Schematic diagram of the 1\,PPS handling component.  Internal components are marked by rectangles.  Pentagons represent input and output signals.  Internal signals are represented by arrows.}
\label{backend_ppscontrol.fig}
\end{center}
\end{figure*}

The 1\,PPS signal from the GPS is crucial because it allows the received sky signal to be matched to the antenna pointing and diagnostic signals.  The C-BASS GPS system suffered from strong $60\,\mathrm{Hz}$ pickup on the 1\,PPS line, which proved difficult to eliminate, and was strong enough to put significant jitter on to the 1\,PPS rising edge. A multistep algorithm was developed which robustly detected the 1\,PPS rising edges and took a running mean of the number of FPGA clock cycles per GPS second through a damping filter.  Any discrepancies between internal timekeeping and the GPS signal were iteratively resolved, providing a stable time reference, accurate to $2\,\mathrm{\upmu s}$. This algorithm is contained within the FPGA firmware in a component called \texttt{pps\_control}.  A schematic illustration of \texttt{pps\_control} and its subcomponents can be found in Fig.~\ref{backend_ppscontrol.fig}.  \texttt{pps\_control} has five subcomponents:
\begin{description}
\item[\texttt{pps\_arrive}] Awaits the arrival of the next 1\,PPS rising edge.  The rising edge is only accepted if the 1\,PPS line has been low for at least 500 clock cycles prior to the rising edge and stays high for at least 500 clock cycles following the rising edge.  If these conditions are met, then an \emph{arrival pulse} is sent to \texttt{pps\_measure}.
\item[\texttt{pps\_measure}] Upon receipt of the \emph{arrival pulse}, calculates the length of the previous second (in clock cycles) taking into account any phase difference between the 1\,PPS and internal counter, as reported by \texttt{pps\_measure}.  The measured second \emph{length} and \emph{phase} are sent to \texttt{pps\_calc\_control} for use in predicting the \emph{length} of future seconds.
\item[\texttt{pps\_calc\_control}] Calculates the running mean of the last thirty-two 1-second lengths to predict the length of the next second in clock cycles.  The predicted second \emph{length} and corrective \emph{phase} shift (which is the measured \emph{phase} divided by 32 to avoid sudden discontinuities) is sent to \texttt{pps\_count} for use during the next second.
\item[\texttt{pps\_count}] Maintains an internal counter representing the best guess of the current time between 1\,PPS arrivals.  The counter counts up to the \emph{predicted second length}, modified by the predicted second \emph{phase}, as calculated by \texttt{pps\_calc\_control}.  This counter is used by \texttt{pps\_measure} to measure the current second's length, as well as to drive several \emph{timing pulses} for use in the firmware DSP chain.
\item[\texttt{pps\_eval}] Ensures the health of the \emph{counter and timing pulses} coming from \texttt{pps\_count} in comparison to the 1\,PPS signal itself.  If all appears to be well, the \texttt{pps\_valid} flag is raised.
\end{description}
These five components work together in a negative feedback loop which is robust to noise on the 1\,PPS line and quickly damps any errors resulting from rare failures to detect the 1\,PPS rising edge.  Indeed, the system is designed to work using predicted second lengths for eight  seconds in the event that no 1\,PPS rising edge is detected. This is sufficient for normal operation.

\begin{figure}
\begin{center}
\includegraphics[width=\columnwidth]{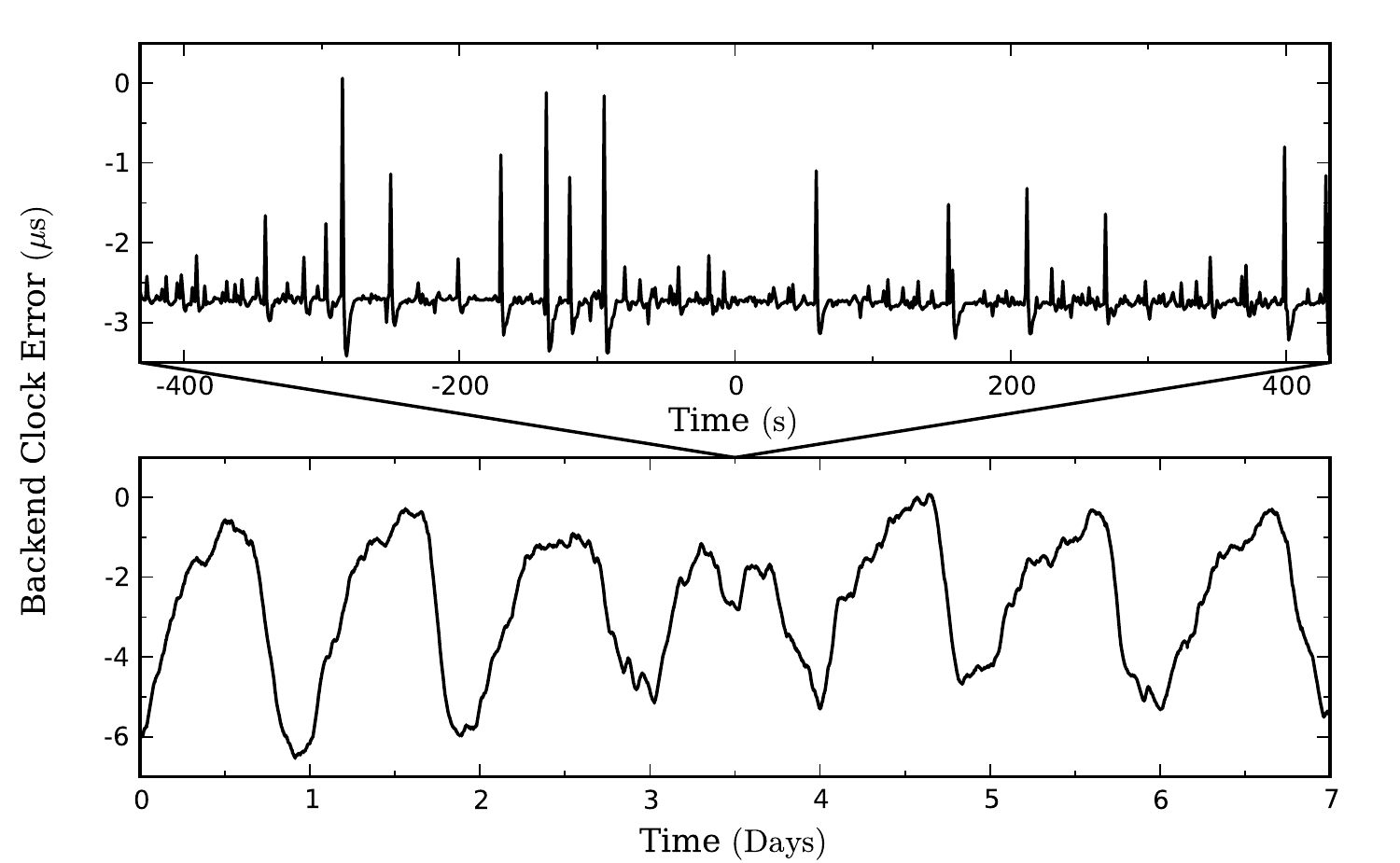}
\caption{Timing errors in detecting the 1\,PPS rising edge.  Lower panel are data from 2012 January 13--19, smoothed over $64$\,s.  Upper panel is zoomed in, without smoothing.}
\label{backend_secondlengths.fig}
\end{center}
\end{figure}

The \texttt{pps\_valid} flag is high when all is working well, and is used by the rest of the firmware to ensure proper timing.  The \emph{predicted second lengths} are bundled with the astronomical data product and are saved in the C-BASS data archive.  A sample of these is shown in Fig.~\ref{backend_secondlengths.fig}.  It is clear from these time series that the timing varies over the day, probably because of temperature variations of the backend hardware.  Errors in the detection of the 1\,PPS rising edge occur on 1-second timescales, but result in timing errors of less than $2\,\mathrm{\upmu s}$ and are quickly damped. These errors are small compared with the integration time and have negligible effect on the astronomical data.

\subsection{Control Signals}\label{backend_firmware_control.sec}

The digital backend controls the digital signals used by the analogue receiver.  These are the ADC triggers, the noise diode control and the six phase switch modulation signals. 

Controlling the ADCs consists of providing them with an FPGA-synchronized clock, sending an acquisition pulse when data are desired, and waiting for the serialized data to arrive.  These jobs are handled by the \texttt{acquisition\_control\_filt} component, which is enabled by the \texttt{adc\_read} register.

The noise diode injects thermal noise into the analogue sky channels prior to combination with the temperature-controlled loads.  This is an essential component of the calibration strategy, and is used to measure the gain and balance on timescales of minutes and longer.  When the \texttt{noise\_on} register is raised, the signal is forwarded to the noise diode immediately and the state of the noise diode is packaged with the backend data products.  

There are two phase switches each for the $I_1$, $I_2$, and $Q, U$ analogue chains (see Fig.~\ref{system-diag}).  Each pair provides four possible phase switch states: 00, 01, 10, and 11.  Although the 00 and 11 states nominally give the same output, as do 01 and 10, using all four states cancels out the effects of unequal complex gains in the signal chains.  Each switch is modulated with a square wave, with each phase-switch pair being displaced by a quarter period.  In this way, the output modulation varies between the four states at the individual phase-switch frequency, and between the degenerate states at twice that rate.  The default individual phase switch frequency is $500\,\mathrm{Hz}$, giving a modulation frequency of $1\,\mathrm{kHz}$.  The $Q$ and $U$ modulation phases are shifted by $180\degr$ relative to the $I_1$ and $I_2$ modulation phase.
The signals are generated in the backend component \texttt{rom\_walsh}.
The same functions are used for demodulation, with a hard-coded delay of $0.98\,\mathrm{\upmu s}$ to account for the latency of the data flow through the system.

\subsection{DSP Chain}\label{backend_dsp.sec}

The signal processing is split into two parallel chains, one for filtered and one for unfiltered data (see Section~\ref{sec:backend-architecture}).  The filtered chain is designed to give maximal rejection of systematic noise while preserving the astronomical signal.  The unfiltered data allows some systematic noise to pass, but preserves absolute power level information.

\begin{figure}
\begin{center}
\includegraphics[width=\columnwidth]{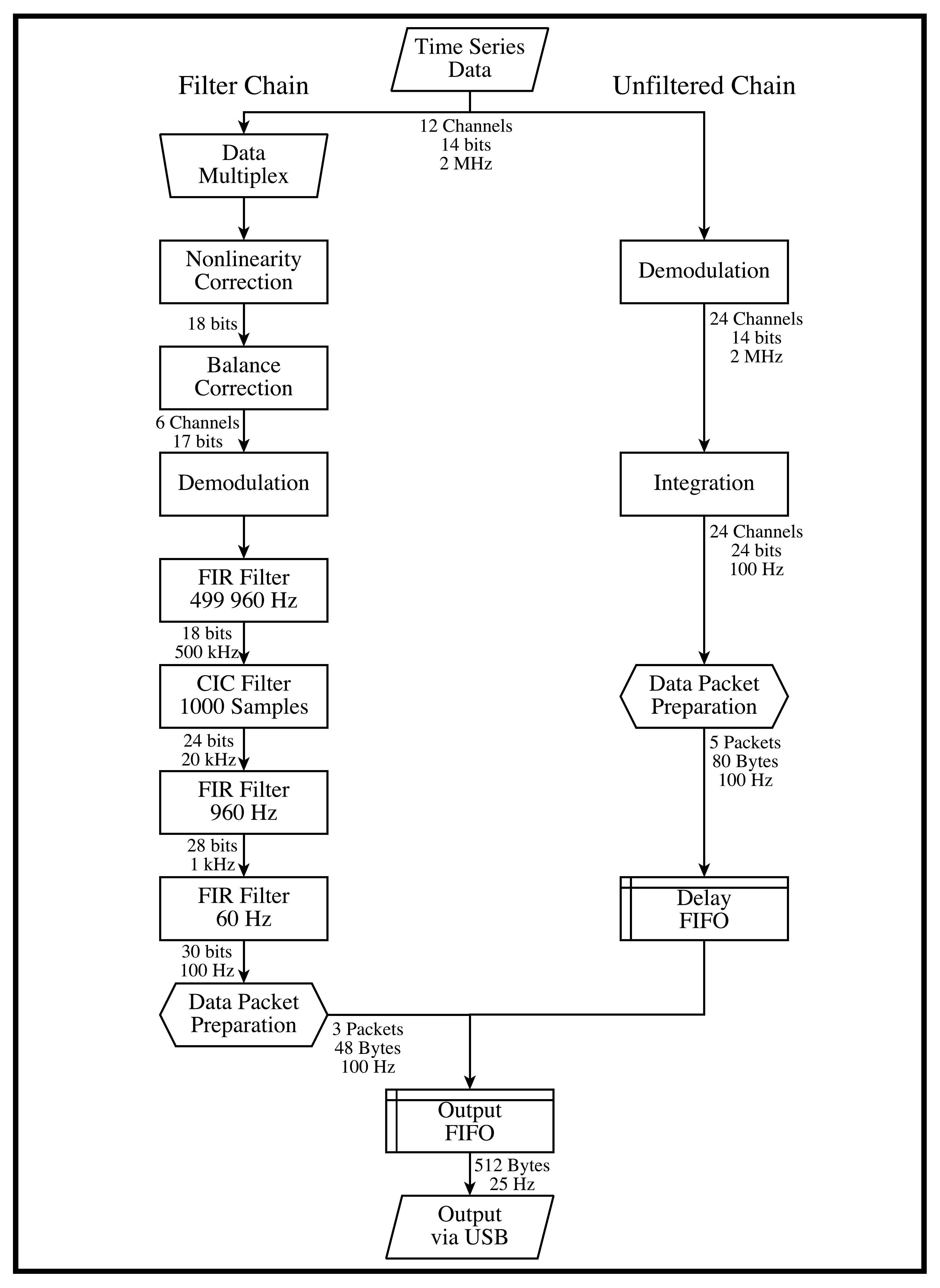}
\caption{Flowchart of the continuous mode DSP, showing data bit-widths and rates at various stages.}
\label{backend_filterchains.fig}
\end{center}
\end{figure}

As described in Section~\ref{backend_hardware.sec}, 14-bit data enter the FPGA via serial transfers from each ADC.  The first step in processing the data is to convert the data from serial to parallel, which is done in the component \texttt{convert\_control}.  The data then enter the \texttt{dsp\_control} component through the \texttt{data\_mux}, which can also be switched to provide simulated data instead.  The unfiltered chain splits the data into 24 channels, representing the two phase-states of each of the 12 analog inputs, while the filtered chain pairwise-subtracts and demodulates the data down to 6 channels.  The unfiltered chain then integrates the data directly, while the filtered chain filters and downsamples the data in four stages, to achieve the final data rate of 100~Hz.  Finally, the outputs of the two chains are packaged together and prepared for USB transfer.  A flowchart of these chains is shown in Fig.~\ref{backend_filterchains.fig}.

The data precision throughout the DSP chain is limited by the FPGA resources.  More bits for a given data point allow higher precision, but at the cost of greater resource requirements.  The FPGA's multipliers and random access memory blocks (BRAM) operate with 18-bit data; using more than 18 bits thus requires double the number of multipliers or RAM blocks.  At the same time, truncating the number of bits can introduce roundoff noise \citep{Rabiner:1975p695}.  Care must be taken to ensure that the roundoff noise is less than the intrinsic noise of the data, to ensure that the final signal-to-noise ratio is not compromised.

\subsubsection{Nonlinearity correction}

The integration and differencing that occurs in the filtered data chain precludes the possibility of doing certain kinds of data corrections off-line. These must therefore be done in the real-time system. In particular, any nonlinearity effects due to the detector diode response, as well as any multiplicative corrections required to correct receiver balance, must be done in real time.  The filter chain allows for these corrections in the \texttt{filter\_condition} component, which applies pre-set nonlinearity and receiver balance corrections to the data.

\begin{figure}
\begin{center}
\includegraphics[width=\columnwidth]{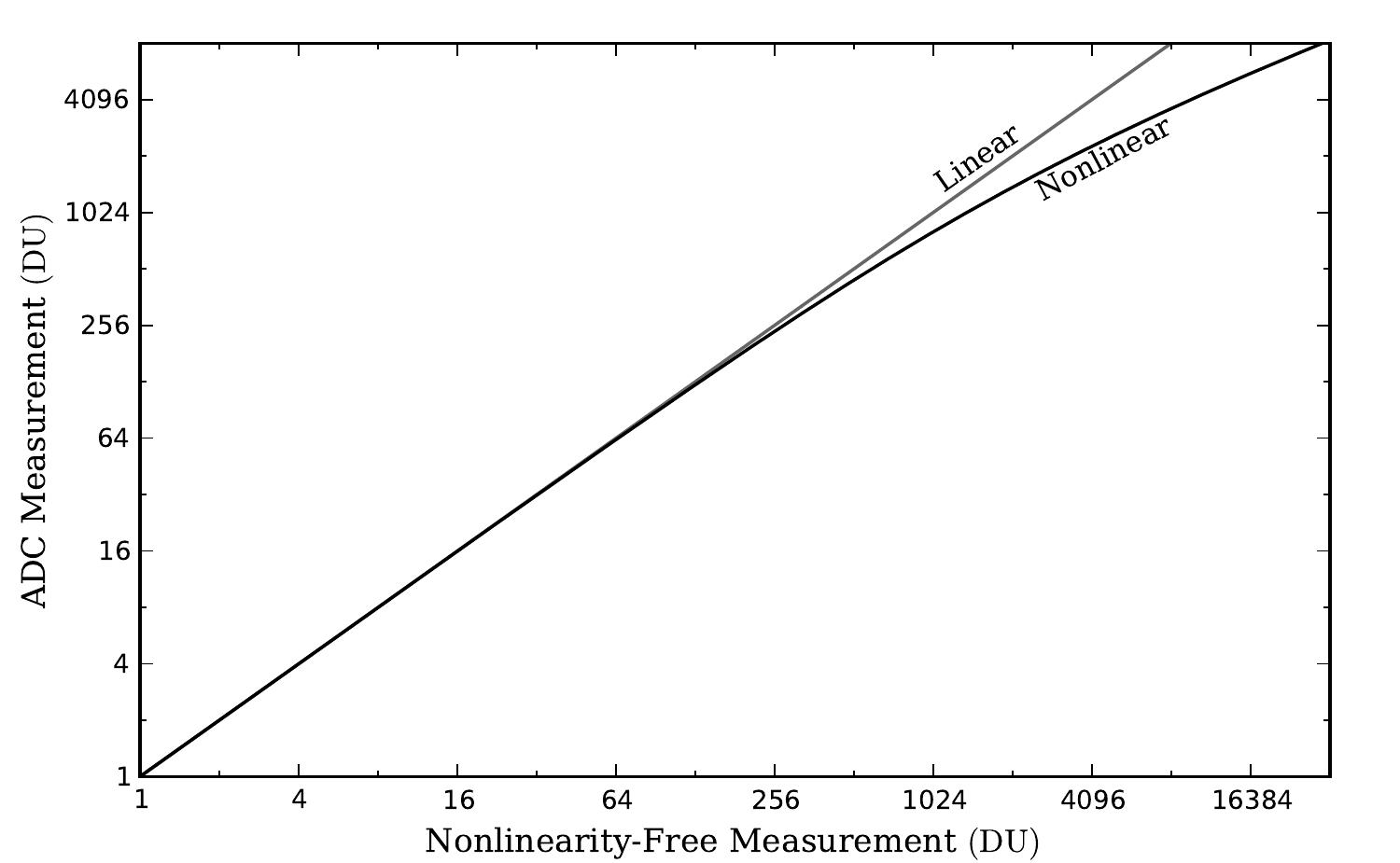}
\caption{Sample nonlinearity curve for the detector diode of Channel~1.  A linear response is shown for comparison.}
\label{backend_nonlinearitycurve.fig}
\end{center}
\end{figure}

Detector nonlinearity is modelled as a quadratic function:
\begin{equation}\label{backend_nonlin_1.eq}
P_\mathrm{NL} = n_1 P_\mathrm{ADC}^2 + n_2 P_\mathrm{ADC} = \left(n_1 P_\mathrm{ADC} + n_2\right)P_\mathrm{ADC}
\end{equation}
where $P_\mathrm{ADC}$ is the measured power and $P_\mathrm{NL}$ the power after nonlinearity correction.  These coefficients were measured for the individual detector diodes in the lab. Typical values are a few times $10^{-4}$ for $n_1$ and between 0.95 and 1.05 for $n_2$. An example curve is shown in Fig.~\ref{backend_nonlinearitycurve.fig}.  On-sky tests demonstrated that in normal astronomical observations nonlinearity was not  detectable, as the diodes were typically operated at relatively low power levels.  

In the case where a nonlinearity correction is required, care must be taken to ensure that the dynamic ranges of the data channels are preserved.  Because the arithmetic is fixed-point rather than floating-point, poor choice of coefficients can lead to under- or over-flow of significant bits.  The \hbox{14-bit} $P_\mathrm{ADC}$ inputs are made 18-bit via zero-padding  the least significant bits (equivalent to multiplying by $2^4$).  Being two's complement, the maximum values are then $2^{17}-2^4$.  The maximum acceptable value of $P_\mathrm{NL}$ is $2^{17}-1$.  The value of $n_1$ should be  maximized to guarantee maximum dynamic range, but be no greater than $2^{17}-1$.  These requirements are satisfied by re-normalizing the nonlinearity coefficients via:
\begin{equation}
N_1 =  2^{17}\, x\, n_1
\end{equation}
and
\begin{equation}
N_2 = 2^4 \, x\,  n_2
\end{equation}
where the multiplicative constant
\begin{equation}
x = \min\left[\frac{1}{2^4}\frac{2^{17}-1}{n_1\left(2^{13}-1\right)+n_2},\frac{1}{2^{17}}\frac{2^{17}-1}{n_1}\right]
\end{equation}
ensures the conditions above.  A single re-normalization factor $x$ is used across all channels; the optimal value can be found by calculating $x$ for each detector diode and taking the minimum of these.  
Equation~\ref{backend_nonlin_1.eq} is then implemented on the backend as
\begin{equation}
P_\mathrm{NL} = \frac{1}{2^{17}}\left[\frac{N_1 \times \left(2^4 P_\mathrm{ADC}\right)}{2^{17}} + N_2\right]\times\left(2^4P_\mathrm{ADC}\right)
\end{equation}
where the multiplications denoted by ``$\times$'' are calculated via multiplexing through a single $18\times18$-bit multiplier and truncating the output to $18$\,bits.

\subsubsection{Receiver balance correction}

Total intensity receiver balance is modelled by two parameters, $\alpha$ and $r$, the first describing the gain imbalance between the paired, total-intensity channels, the second describing the mismatch between sky and reference load temperatures \citep{2015MNRAS.446.1252K}. Both of these represent linear operations on the paired channels.  Owing to the limited resources of the FPGA, they were combined into a single step:
\begin{equation}\label{backend_alpha_1.eq}
P_\mathrm{sky-load} = \left[\frac{1+r}{\alpha} + \left(1-r\right)\right]P_\mathrm{NL,A} - \left[\frac{1+r}{\alpha} - \left(1-r\right)\right]P_\mathrm{NL,B}
\end{equation}
where $P_\mathrm{sky-load}$ is the balance-corrected difference between sky and reference load power, and  $P_\mathrm{NL,A}$ and $P_\mathrm{NL,B}$ are the powers measured from the paired channels (with nonlinearity correction optionally applied). 

As with the nonlinearity correction, the coefficients need to be carefully normalized.  The constraints are similar to those above, though in this case it is desired that $P_\mathrm{sky-load}$ will be $17\,\mathrm{bit}$ and so must be less than $2^{16}-1$.  This is guaranteed by using
\begin{equation}
R_A = x\left[\frac{1+r}{\alpha} + \left(1-r\right)\right]
\end{equation}
and
\begin{equation}
R_B = x\left[\frac{1+r}{\alpha} - \left(1-r\right)\right]
\end{equation}
where
\begin{equation}
x = \frac{2^{16}}{(1+r)/\alpha + \left|1-r\right|}
\end{equation}
The backend then applies Equation~\ref{backend_alpha_1.eq} through
\begin{equation}
P_\mathrm{sky-load} = \frac{1}{2^{17}}\left(R_A\times P_\mathrm{NL,A} - R_B\times P_\mathrm{NL,B}\right)
\end{equation}
where, again, the multiplications are multiplexed through a single multiplier, and truncated to $17$~bits.

The $\alpha$ correction acts only as a multiplicative offset if the $r$ correction is not applied (i.e., $r=1$), and the need for an $r$ correction can be mitigated by carefully matching the sky and load temperatures using the procedures described in another  paper (Muchovej et al., in preparation). The sky and load temperatures were balanced for most of the C-BASS survey observations, so no correction was applied in the backend (i.e., $\alpha=0$ and $r=1$ were used).  

\subsubsection{Demodulation}

In the absence of modulation, astronomical signals appear in the time series at frequencies below $40\,\mathrm{Hz}$. 
With modulation, the astronomical signal is mixed to the phase-switch frequency and its harmonics.  Demodulation brings the astronomical signal back down below $40\,\mathrm{Hz}$, but it also shifts any low-frequency systematic noise that enters the data post-modulation  up to the modulation frequency.  The systematic noise can then be rejected by low-pass filtering.

To accomplish the demodulation, the modulating signal that is generated in the backend is delayed by the hardware latency (described in Section~\ref{backend_firmware_control.sec}) and any firmware latency experienced by the time-series data up to this point.  Thus synchronized, the pairwise subtracted data have their signs flipped when in modulation state 01 or 10.  The six demodulated, $17$-bit data are then ready to be filtered.

Demodulation is also performed in the unfiltered chain.  In this case, however, the individual modulation states are preserved and no pairwise subtraction takes place.  Thus the unfiltered chain maintains 24 data channels through to integration.

\subsubsection{Filtering and Integration}\label{backend_filtering.sec}

The final data rate of $100\,\mathrm{Hz}$ requires frequencies above $50\,\mathrm{Hz}$ to be rejected.  A single-stage filter capable of meeting this goal for $2\,\mathrm{MHz}$ data would require more resources than were available in the FPGA, so it was  necessary to perform this filtering in four stages.  After each filter, the data are downsampled, allowing for greater efficiency in multiplier utilization for those filters that follow.  

The filters are all of the finite impulse response (FIR) variety, meaning that the filtered data are linear combinations of a finite number of input samples.  Three of the filters (the first, third, and fourth) are simple, decimating, low-pass filters with predefined filter coefficients.  The second filter is a cascaded integrator-comb (CIC) filter \citep{CIC} with fractional downsampling.  The choice of filters was motivated by the necessary fixed-point precision and the multiplier and memory resources required by the various types of filters.  A further consideration was aliasing:  after each filter and decimation, it is essential that any frequencies aliased between 0 and $40\,\mathrm{Hz}$ be sufficiently attenuated.

The CIC filter had the largest influence on the overall design. The CIC filter is a highly efficient implementation of a low-pass filter with decimation, which requires only additions rather than multiplications. It has limited scope for defining the filter characteristics (e.g., roll-off sharpness), but these can be improved by the other FIR filters in the DSP chain. The fractional downsampling required by this filter is the interface between the digital backend $50\,\mathrm{MHz}$ clock and the GPS-supplied 1\,PPS clock.  Time-series data enter this filter on the former clock, and exit synchronized to the latter.  However, this fractional downsampling relies upon downsampling signals originating in the \texttt{pps\_control} component, which requires a custom design for the filter.  Such a filter operates by keeping a running mean of the data time series, with newer data points being added and older data points being subtracted.  This requires keeping track of the older data points, which is done through FIFO memory.  A single block of RAM on this FPGA can only hold 1024 $18$-bit values.  BRAM usage is thus optimized by placing this filter early enough in the chain that $18$ bits are sufficient, late enough that a length of 1000 samples is adequate, and followed by strong enough filtering that a sharp cutoff at $50\,\textrm{Hz}$ is still achievable.

The resulting design is that shown in Fig.~\ref{backend_filterchains.fig}, with the filter properties listed in Table~\ref{backend_filters.tab}.  The filters are characterized by their passband frequencies $f_\textrm{pass}$, stopband frequencies $f_\textrm{out}$, the passband ripple $A_\mathrm{pass}$, and the stopband attenuation $A_\mathrm{stop}$.  The input frequencies, filter orders, decimation rates, and output bitwidths are also listed.  

\input{Tables/backend_filters.tex}

The first filter was designed under the limitations of using a single multiplier/BRAM pair and providing the CIC filter with $18$-bit inputs.  This filter decimates by a factor of four, has a passband below $40\,\mathrm{Hz}$, and a stopband above $499\,960\,\mathrm{Hz}$ (with $100\,\mathrm{dB}$ of rejection).  This is a rather broad filter, only requiring $15$ coefficients.  The output  data are at a $500\,\mathrm{kHz}$ rate and $18$-bit resolution.  The CIC filter follows, with a length of $1000$ samples and a decimation factor of 25.  The CIC filter's frequency response is a sinc function, which falls off only as $1/\nu$ but has nulls at multiples of $5\,\mathrm{kHz}$.  The choice of decimation rate thus places those frequencies aliased to the passband at nulls in the CIC response.  Further, any periodic artefacts due to the modulation itself will be nulled as well.  As mentioned above, the downsampling rate is synchronized to the 1\,PPS timing so that the output data have a rate of exactly $20\,\mathrm{kHz}$.  The filter length requires an output precision of $24\,\mathrm{bit}$.  For this filter, each channel requires its own BRAM, and so a total of six on the FPGA are used.

The last two filters decimate by factors of $20$ and $10$ respectively to reach the final data rate of $100\,\mathrm{Hz}$.  Achieving the necessary stopbands at $940$ and $60\,\mathrm{Hz}$, with attenuations of $100$ and $80\,\mathrm{dB}$, required filter lengths of 119 and 239, respectively.  Output precisions of $28\,\mathrm{bit}$ and $30\,\mathrm{bit}$ were used.  These filters were achieved on the FPGA using 3 multipliers plus 4 blocks of RAM for the first, and 3 plus 6 for the second.  Altogether, the filter chain used 17 of the FPGA's 24 multiplier/BRAM pairs.

As mentioned in Section~\ref{backend_firmware_control.sec}, only a small number of modulation frequencies are compatible with this filter chain.  There are two constraints.  First, the modulation period must be divisible by $16$ to accommodate the four phase states and the first decimation by four.  Second, the CIC period must be divisible by one quarter of the modulation period in order to guarantee removal of modulation artefacts.  Eight frequencies meet these requirements and are listed in Table~\ref{backend_modulationfrequencies.tab}.  The default was chosen as the lowest of these in order to minimize attenuation from the detector diode video bandwidths.

\input{Tables/backend_modulationfrequencies.tex}

\begin{figure}
\begin{center}
\includegraphics[width=\columnwidth]{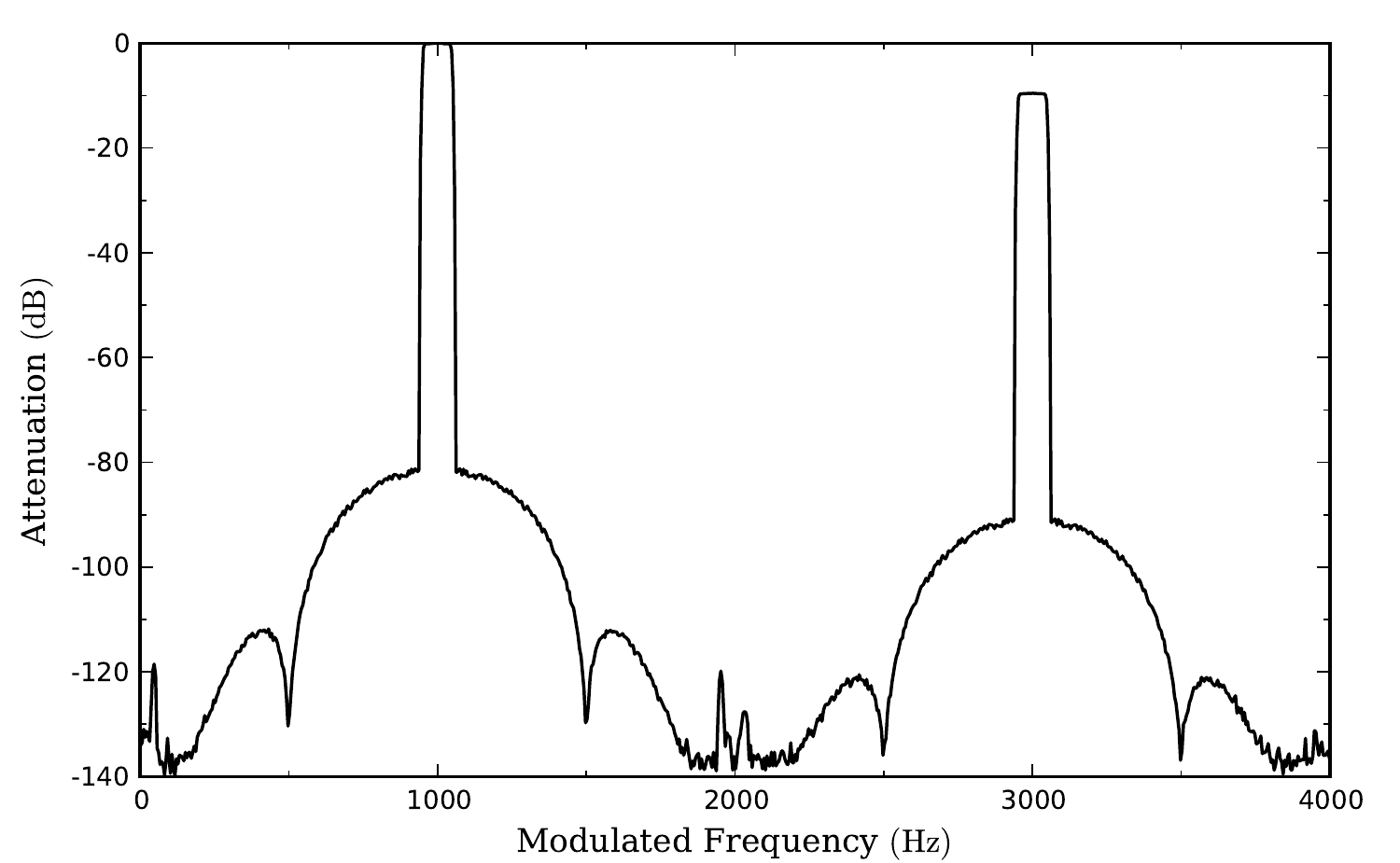}
\caption{Frequency response of the full DSP chain.  Modulated frequency refers to the frequency prior to demodulation.  The passbands are centred upon the modulation frequency and its harmonics, where the sky signal appears.}
\label{backend_filterattenuation.fig}
\end{center}
\end{figure}

Design and simulation of the full filter chain was performed using the \texttt{MATLAB} software suite.\footnote{The MathWorks, Inc.\ \url{https://www.mathworks.com}}  This software allowed for the easy calculation of filter coefficients given a wide range of desired filter properties, while allowing for the full filter chain (including effects of bit precision) to be directly simulated and evaluated.  The full filter frequency response (in modulated frequency space) is shown in Fig.~\ref{backend_filterattenuation.fig}.  The modulated sky signal appears at $1\,\mathrm{kHz}$ modulation frequency and its odd harmonics (3, 5, 7...kHz).  Any spurious signals outside this range are attenuated at greater than $80\,\mathrm{dB}$.

\begin{figure*}
\begin{center}
\includegraphics[width=0.7\textwidth]{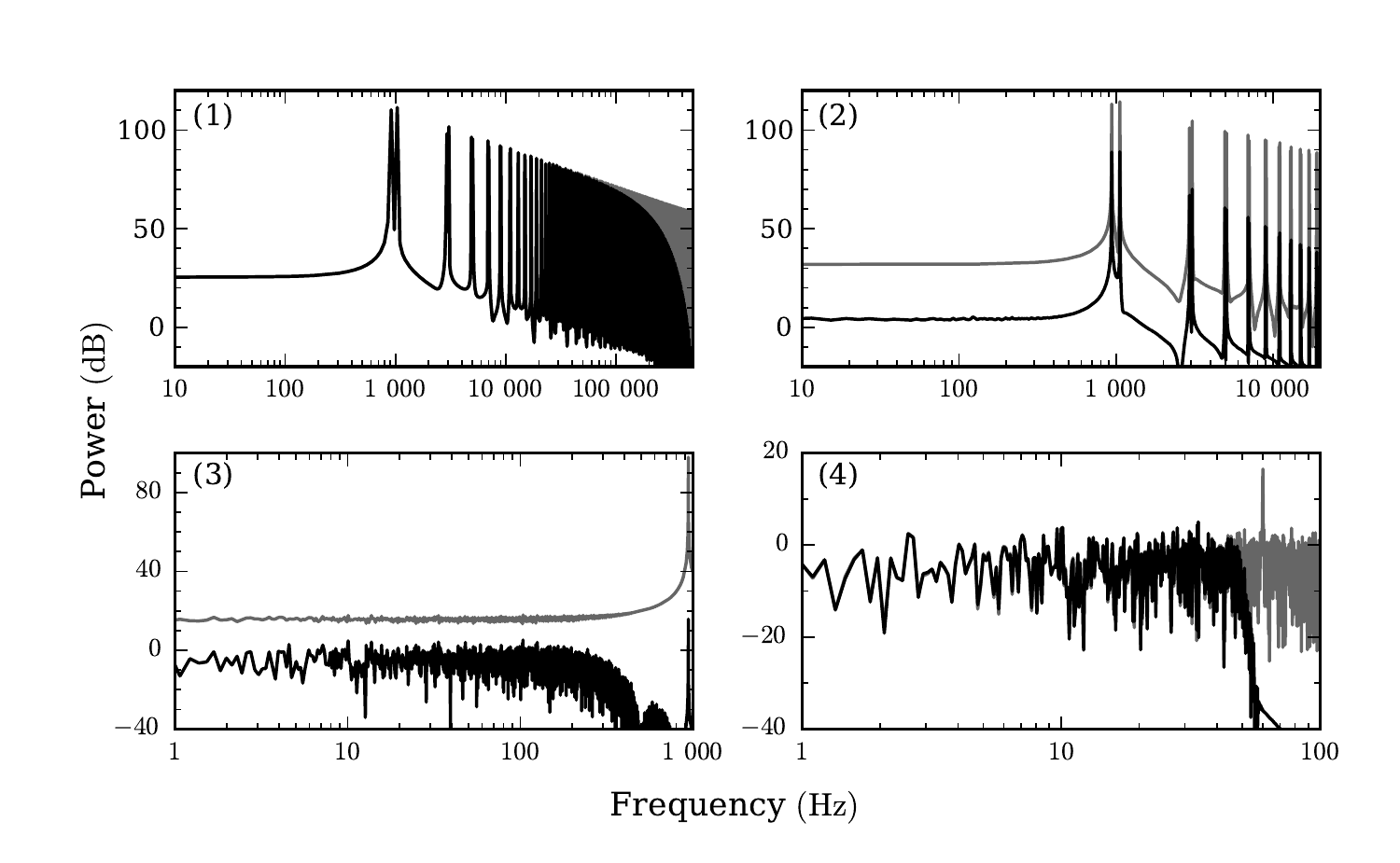}
\caption{Frequency responses for the four filters (1, 2, 3, 4) to a simulated data stream containing strong 60 Hz contamination.  The $60\,\mathrm{Hz}$ pickup has been mixed up to kilohertz frequencies by the demodulation.  \textit{Grey}: data spectrum prior to filter; \textit{black}: after filter. Filter numbers correspond to the descriptions in Table \ref{backend_filters.tab}. }
\label{backend_filtersimulation.fig}
\end{center}
\end{figure*}

The performance was further evaluated with a second simulation, shown in Fig.~\ref{backend_filtersimulation.fig}.  Using an input time series with noise properties similar to that of the C-BASS receiver and a $60\,\mathrm{Hz}$ signal with an amplitude $\sim10\,000$ greater than normal, the spectrum before and after each filter was calculated.  The rejection of the interfering signal is dramatic.  Effects of intermediate bit-precision between the individual filters were tested using this simulation, confirming that the precisions chosen increased the noise of the output time series by less than $1\,\%$.

\begin{figure}
\begin{center}
\includegraphics[width=\columnwidth]{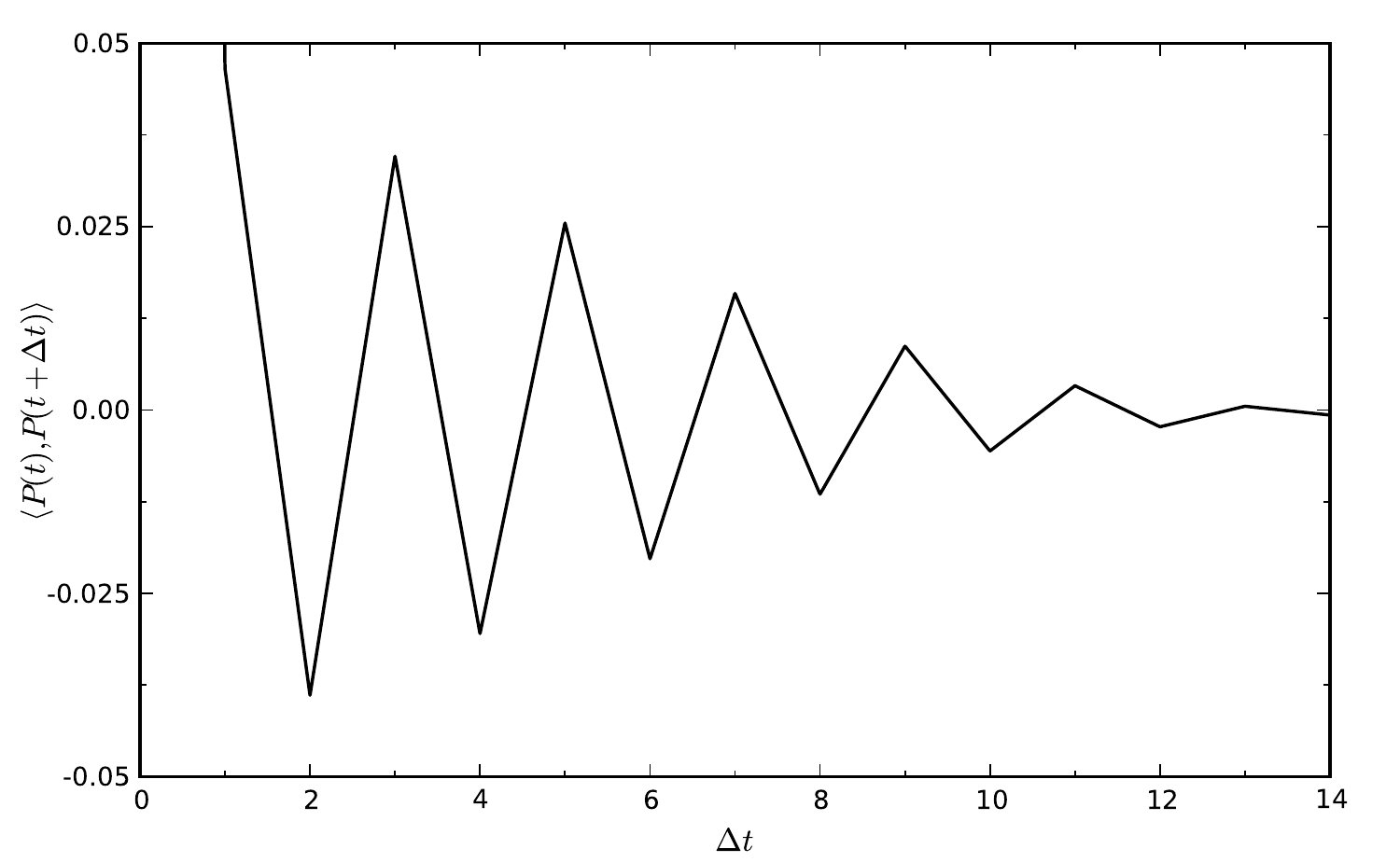}
\caption{Autocorrelation function of the filter chain, normalized to 1 at $\Delta t=0$.  Note that $\Delta t$ is in units of $10\,\mathrm{ms}$ samples.}
\label{backend_autocorrelation.fig}
\end{center}
\end{figure}

\begin{figure}
\begin{center}
\includegraphics[width=\columnwidth]{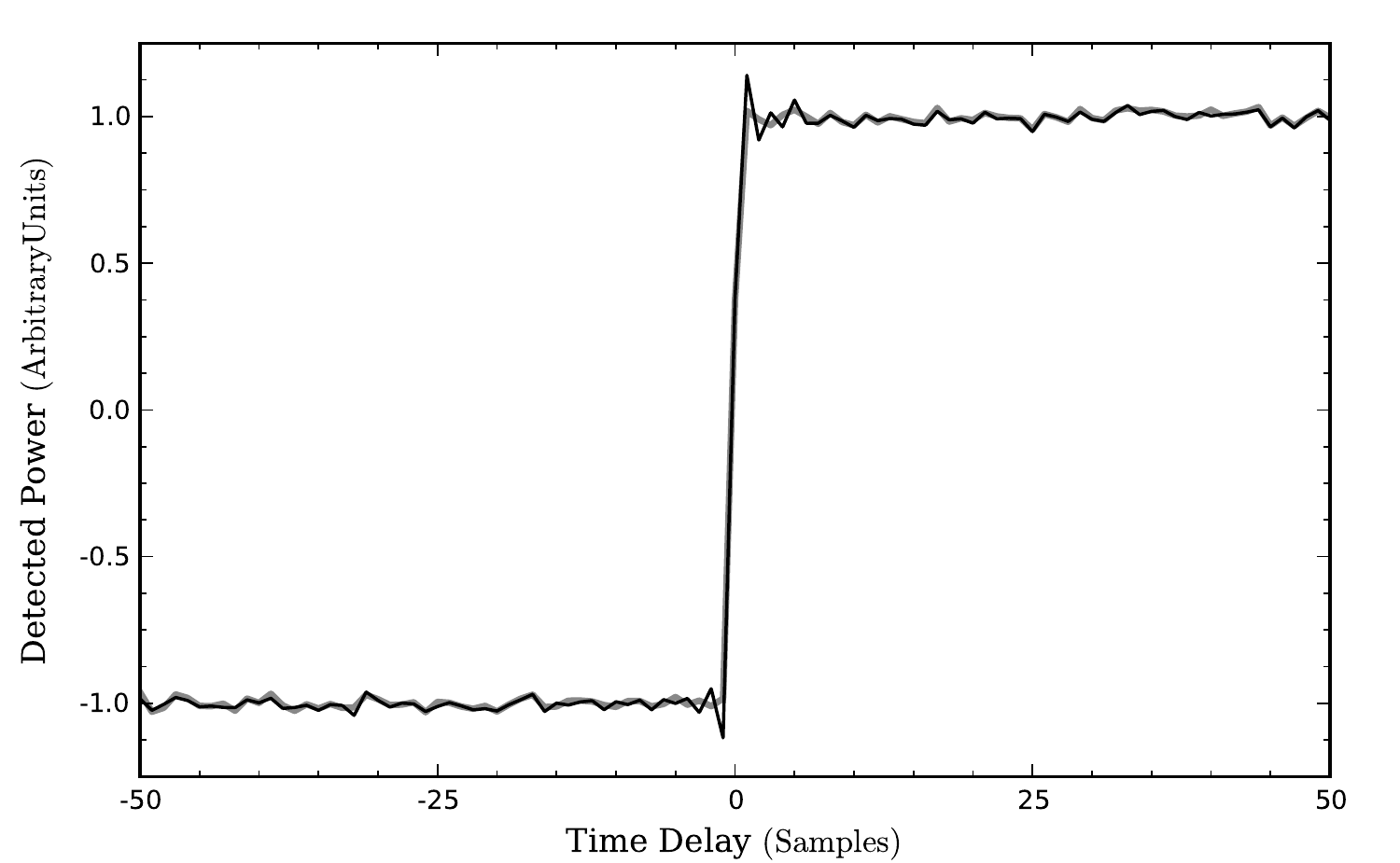}
\caption{Effect of the filter autocorrelation function.  At sample 0, the noise diode is switched on. The filter induces ringing around the transition  (\textit{black}) that is not present in the unfiltered data (\textit{grey}). }
\label{backend_filterringing.fig}
\end{center}
\end{figure}

This excellent frequency-domain response inevitably comes at the expense of some time-domain correlation.  The effective filter length spans twenty-nine $10\,\mathrm{ms}$ samples, so the filtered data are autocorrelated across this timescale.  This autocorrelation is a property of the filter chain and will manifest most strongly as oscillations preceding and following step functions (such as when the noise diode is turned on or off).  Analytical calculation of the autocorrelation function is precluded by the fractional downsampling in the CIC filter, so it was calculated numerically instead.  This calculation was performed by injecting white, Gaussian noise into the simulated filter chain and finding the autocorrelation of the result.  The function is plotted in Fig.~\ref{backend_autocorrelation.fig} and an example of the effect on a step function in real C-BASS data is shown in Fig.~\ref{backend_filterringing.fig}. In C-BASS survey data, the effect is to smear the beam very slightly in the scan direction. However, the small level of the autocorrelation, and its rapid decay, mean that the actual distortion to the beam in final science data (when further smoothing is applied to the maps) is much less than one part in $10^3$.

The unfiltered processing chain uses an integration scheme with fractional triggering, similar in concept to the CIC filter.  In this case, the idea is to accumulate $10^4$ samples for each of the 24 demodulated channels, with the accumulations being triggered by the 1\,PPS-synchronized counter contained in \texttt{pps\_control}.  The resulting time series are produced at exactly $100\,\mathrm{Hz}$.  An output precision of $24\,\mathrm{bit}$ was chosen, which is less than that from the filters.  Although concern about truncation noise is reasonable, this precision is constrained by the backend output bandwidth.  Because these data are used for receiver balance monitoring and other diagnostic tests rather than for science data products, the concession in output precision is not expected to adversely affect the final C-BASS data products.

The demodulation of the data into 24 channels leaves these data susceptible to systematic effects during the switching of modulation states.  The filtered chain is designed to mitigate this problem; this does not happen in the unfiltered chain.  Optional pre-integration flagging of affected data is allowed instead.  A constant number of data points can be flagged and trimmed following each phase state transition.

Although the individual integrations are ostensibly independent, the mismatch between backend and GPS timing implies that a few integrations per second will overlap by a single $2\,\mathrm{MHz}$ data point. However, the  maximum correlation between adjacent integrations due to this effect is at the $0.01\,\%$ level and is thus unimportant.

Compared to the filtered chain, the unfiltered integrations are calculated with a much shorter latency, so the integrated time series needs to be stored prior to packaging with the filtered data. This storage requires the use of a single block of RAM.  The full DSP chain therefore uses $75\,\%$ of the FPGA's multiplier and RAM resources.

\begin{figure}
\begin{center}
\includegraphics[width=\columnwidth]{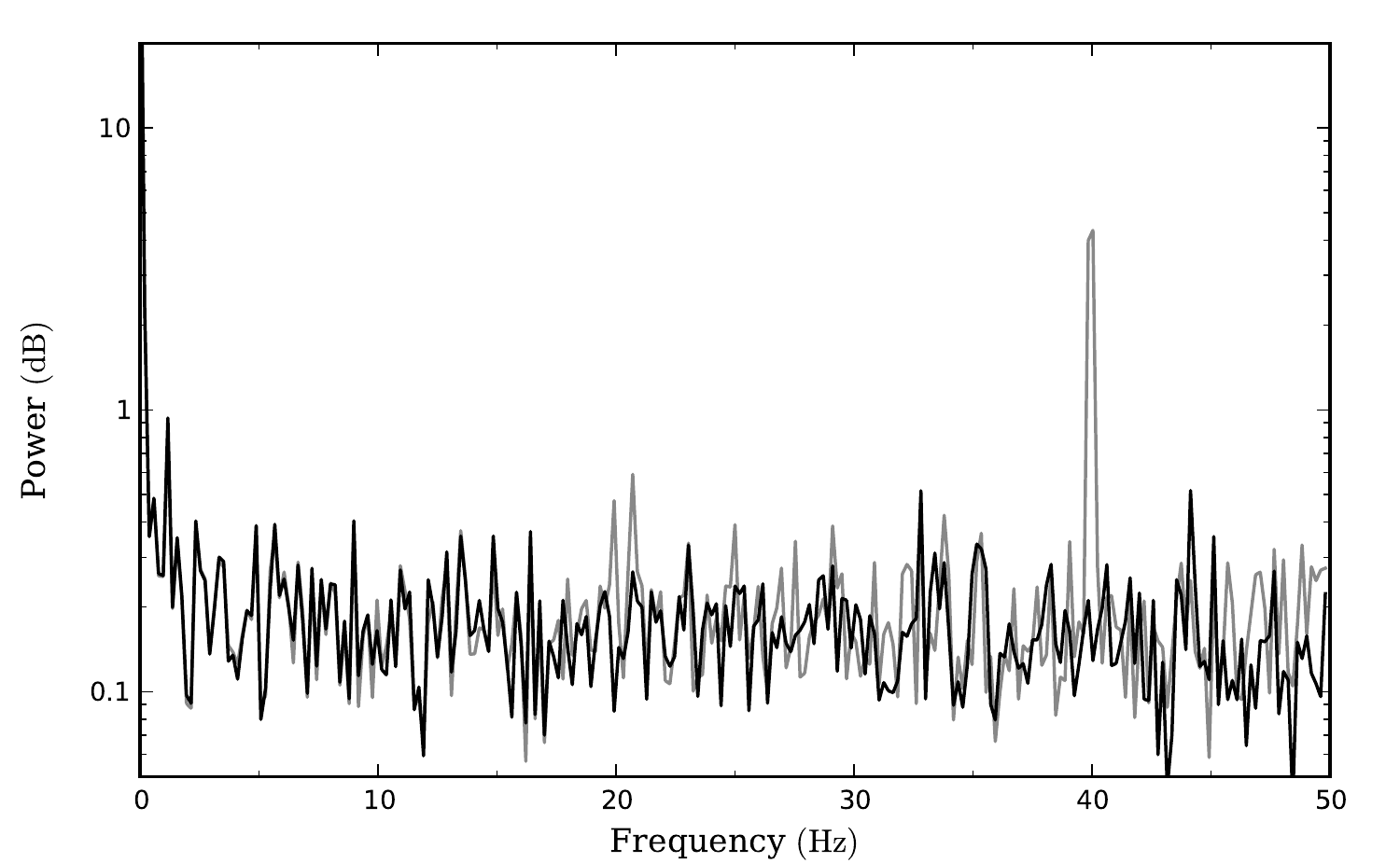}
\caption{Power spectra of filtered (\textit{black}) and unfiltered (\textit{grey}) $I_1$ data from 2012 January 15.  $60\,\mathrm{Hz}$ pickup and harmonics appear aliased at $20\,\mathrm{Hz}$ and $40\,\mathrm{Hz}$ without filtering.}
\label{backend_filter_vs_unfilter_ps.fig}
\end{center}
\end{figure}

Fig.~\ref{backend_filter_vs_unfilter_ps.fig} shows the power spectra of a short stretch of real C-BASS data, processed through the filtered and unfiltered chains. The reduction in aliased $60\,\mathrm{Hz}$ pickup at $20\,\mathrm{Hz}$ and $40\,\mathrm{Hz}$ is clear.

\subsection{USB Microcontroller}

Communication between the digital backend and the control system computer is provided through the USB microcontroller.  This chip is built around an industry-standard 8051 microprocessor which can be programmed in the C language.  Programming was done using the \texttt{Kiel} programming environment, which produced a compiled, hexadecimal file, sent to the microcontroller from the control computer upon startup of the backend.  The microcontroller contains no non-volatile memory, so this programming file must be sent to the backend after every power cycle.

The USB firmware was provided by the LiCAS project, with only superficial modifications needed to accommodate the C-BASS backend.  A new C driver was written to allow control from a UNIX platform (the LiCAS system being run from the Windows operating system) and was incorporated into the telescope control system.  This driver worked with the microcontroller firmware to define a number of low-level control commands.   The commands are interpreted by the microcontroller firmware and sent to the primary FPGA via its parallel, 8-bit \texttt{pc} and \texttt{pe} inputs.

The USB microcontroller is clocked at $25\,\mathrm{MHz}$, with the clock signal being provided by the primary FPGA.  This different clock speed represents a second clock domain within the primary FPGA, which was bridged through a dual-clock FIFO, requiring use of two of the FPGA's blocks of RAM.  After crossing this clock domain interface, the data are sent to the microcontroller via $16\,\mathrm{bit}$ of the FPGA's parallel, $18\,\mathrm{bit}$ output \texttt{usb\_in}.  Once the microcontroller has received all $512\,\mathrm{B}$ for a given bulk transfer, the microcontroller makes those data available to the control system and begins accepting data for the next bulk transfer.  

\section{Conclusions}
\label{sec:conclusions}

We have described the C-BASS digital backend, which implements the readout, phase switch control and demodulation, filtering and integration of the C-BASS radiometer/polarimeter data. It provides for real-time correction of detector nonlinearity and receiver imbalance, and also provides two parallel data streams: one filtered and phase-switched to remove out-of-band signals and to cancel gain imbalances in the receiver, the other preserving the total power and gain imbalance information for diagnostic purposes. A cascaded combination of FIR and CIC filters provides an efficient implementation of the filtering and integration functions. These operations are efficiently packed in to a relatively small capacity FPGA, and data transfers to the telescope control system are handled by a reliable USB interface. The filters suppress 60\,Hz mains pickup by $\sim 80$\,dB, while introducing $\sim 2$ per cent correlations between closely-spaced time samples.   

The backend was used to process the data for the northern part of the C-BASS survey, which were recorded from 2009 December until the receiver was decommissioned in 2015 April \citep{projectpaper}. Some early results from the northern survey have been presented by \citet{2015MNRAS.448.3572I} and \citet{ncppaper}, and the full results will be published in forthcoming papers.

\section*{Acknowledgements}

We thank Martin Shepherd, David Hawkins, Russ Keeney, and Erik Leitch for advice and assistance with the design and construction of the backend. 
The work at the California Institute of Technology and Owens Valley Radio Observatory was supported by National Science Foundation (NSF) awards~AST-0607857, AST-1010024, AST-1212217, and AST-1616227, and by NASA award NNX15AF06G. 
The work at Oxford was supported by funding from STFC, the Royal Society and the University of Oxford.  
CD acknowledges support from an ERC Grant no.\,307209. 
OGK acknowledges the support of a Dorothy Hodgkin Award in funding his studies while a student at Oxford, and the support of a W.M.~Keck Institute for Space Studies Postdoctoral Fellowship at Caltech. 
This work formed part of MAS's PhD thesis \citep{stevenson_thesis}.


\bibliographystyle{mnras}
\bibliography{backend}

\bsp	
\label{lastpage}
\end{document}

%% file: Tables/backend_chips.tex
\begin{table*}
\caption{Components of the C-BASS Digital Backend.}
\label{backend_chips.tab}
\begin{tabular}{llll} 
\hline
Component & Manufacturer & Part Number  & Comment\\ \hline
Serial 14-Bit ADC & Linear Technology &  LTC1403A-1 & $2\,\mathrm{MS/s}$ sampling \\ 
Spartan-3 FPGA & Xilinx & XC3S1000-4 & Main and Secondary \\ 
Synchronous DRAM & Micron Technology & MT48LC32M16A2 & Two installed \\ 
USB Microcontroller & Cypress Semiconductor & CY7C68013A & ``Default'' Configuration \\ 
\hline
\end{tabular}
\end{table*}

%% file: Tables/backend_timescales.tex
\begin{table}
\caption{Time bases of the C-BASS Digital Backend.}
\label{backend_timescales.tab}
\begin{tabular}{lrrr}
\hline
Timescale & Family & Frequency & Period \\ \hline
Backend Clock & Clock & $50\,\mathrm{MHz}$ & $20\,\mathrm{ns}$ \\
USB Clock & Clock & $25\,\mathrm{MHz}$ & $40\,\mathrm{ns}$ \\
ADC Sampling & Clock & $2\,\mathrm{MHz}$ & $500\,\mathrm{ns}$ \\
Modulation & Clock & $1\,\mathrm{kHz}$ & $1\,\mathrm{ms}$ \\
Integration & 1PPS & $100\,\mathrm{Hz}$ & $10\,\mathrm{ms}$ \\
USB Transfer & 1PPS & $25\,\mathrm{Hz}$ & $40\,\mathrm{ms}$ \\
GPS 1PPS & 1PPS & $1\,\mathrm{Hz}$ & $1\,\mathrm{s}$ \\
\hline
\end{tabular}
\end{table}%

%% file: Tables/backend_filters.tex
\begin{table*}
\caption{Individual Filter Details.}
\label{backend_filters.tab}
\begin{tabular}{ccrrrrrccc}
\hline \noalign{\smallskip}
Filter & Filter & $f_\mathrm{in}$ & $f_\mathrm{pass} $ &  $f_\mathrm{stop}$ & $A_\mathrm{pass}$ &  $A_\mathrm{stop}$ & Order & Decimation & Output \\ 
Number & Type & [kHz] & [Hz] & [Hz] & [dB] & [dB] & & & Bitwidth \\ \hline \noalign{\smallskip}
1 & FIR & $2\,000$ & 40 & $499\,960$ & $10^{-6}$ & 100 & 14 & 4 & 18 \\
2 & CIC & $500$ & $\cdots$ & $\cdots$ & $\cdots$ & $\cdots$ & 1000 & 25 & 24 \\
3 & FIR & $20$ & 40 & 960 & $10^{-4}$ & 100 & 118 & 20 & 28 \\
4 & FIR & $1$ & 40 & 60 & $10^{-4}$ & 80 & 238 & 10 & 30 \\ \noalign{\smallskip}
\hline 
\end{tabular}
\end{table*}

%% file: Tables/backend_modulationfrequencies.tex
\begin{table}
\caption{Allowable Modulation Frequencies for the filter chain.}
\label{backend_modulationfrequencies.tab}
\begin{center}
\begin{tabular}{cc}
\hline \noalign{\smallskip}
Switch Frequency&Modulation Frequency \\
{[kHz]}& [kHz] \\ 
\hline \noalign{\smallskip}
$0.5$ &  1 \\
$1.0$ & 2 \\
$2.5$ & 5 \\
$5.0$ & 10 \\
$12.5$ & 25 \\
$25.0$ & 50 \\
$62.5$ & 125 \\
$125.0$ & 250 \\ \noalign{\smallskip}
\hline
\end{tabular}
\end{center}
\end{table}

%% file: cbass-backend-paper.bbl
\begin{thebibliography}{}
\makeatletter
\relax
\def\mn@urlcharsother{\let\do\@makeother \do\$\do\&\do\#\do\^\do\_\do\%\do\~}
\def\mn@doi{\begingroup\mn@urlcharsother \@ifnextchar [ {\mn@doi@}
  {\mn@doi@[]}}
\def\mn@doi@[#1]#2{\def\@tempa{#1}\ifx\@tempa\@empty \href
  {http://dx.doi.org/#2} {doi:#2}\else \href {http://dx.doi.org/#2} {#1}\fi
  \endgroup}
\def\mn@eprint#1#2{\mn@eprint@#1:#2::\@nil}
\def\mn@eprint@arXiv#1{\href {http://arxiv.org/abs/#1} {{\tt arXiv:#1}}}
\def\mn@eprint@dblp#1{\href {http://dblp.uni-trier.de/rec/bibtex/#1.xml}
  {dblp:#1}}
\def\mn@eprint@#1:#2:#3:#4\@nil{\def\@tempa {#1}\def\@tempb {#2}\def\@tempc
  {#3}\ifx \@tempc \@empty \let \@tempc \@tempb \let \@tempb \@tempa \fi \ifx
  \@tempb \@empty \def\@tempb {arXiv}\fi \@ifundefined
  {mn@eprint@\@tempb}{\@tempb:\@tempc}{\expandafter \expandafter \csname
  mn@eprint@\@tempb\endcsname \expandafter{\@tempc}}}

\bibitem[\protect\citeauthoryear{{Dickinson} et~al.,}{{Dickinson}
  et~al.}{2018}]{ncppaper}
{Dickinson} C.,  et~al., 2018, preprint, \href
  {http://adsabs.harvard.edu/abs/2018arXiv181011681D} {} (\mn@eprint {arXiv}
  {1810.11681})

\bibitem[\protect\citeauthoryear{{Hamaker} \& {Bregman}}{{Hamaker} \&
  {Bregman}}{1996}]{1996A&AS..117..161H}
{Hamaker} J.~P.,  {Bregman} J.~D.,  1996, \aaps, \href
  {http://adsabs.harvard.edu/abs/1996A%26AS..117..161H} {117, 161}

\bibitem[\protect\citeauthoryear{Hogenauer}{Hogenauer}{1981}]{CIC}
Hogenauer E.~B.,  1981, IEEE Transactions on Acoustics, Speech and Signal
  Processing, 29 (2), 155

\bibitem[\protect\citeauthoryear{{Irfan} et~al.,}{{Irfan}
  et~al.}{2015}]{2015MNRAS.448.3572I}
{Irfan} M.~O.,  et~al., 2015, \mn@doi [\mnras] {10.1093/mnras/stv212}, \href
  {http://adsabs.harvard.edu/abs/2015MNRAS.448.3572I} {448, 3572}

\bibitem[\protect\citeauthoryear{{Jarosik} et~al.,}{{Jarosik}
  et~al.}{2003}]{2003ApJS..145..413J}
{Jarosik} N.,  et~al., 2003, \mn@doi [\apjs] {10.1086/346080}, \href
  {http://adsabs.harvard.edu/abs/2003ApJS..145..413J} {145, 413}

\bibitem[\protect\citeauthoryear{{Jones} et~al.,}{{Jones}
  et~al.}{2018}]{projectpaper}
{Jones} M.~E.,  et~al., 2018, \mn@doi [\mnras] {10.1093/mnras/sty1956}, \href
  {http://adsabs.harvard.edu/abs/2018MNRAS.480.3224J} {480, 3224}

\bibitem[\protect\citeauthoryear{{King} et~al.,}{{King}
  et~al.}{2014}]{2014MNRAS.438.2426K}
{King} O.~G.,  et~al., 2014, \mn@doi [\mnras] {10.1093/mnras/stt2359}, \href
  {http://adsabs.harvard.edu/abs/2014MNRAS.438.2426K} {438, 2426}

\bibitem[\protect\citeauthoryear{{King} et~al.,}{{King}
  et~al.}{2015}]{2015MNRAS.446.1252K}
{King} O.~G.,  et~al., 2015, \mn@doi [\mnras] {10.1093/mnras/stu2172}, \href
  {http://adsabs.harvard.edu/abs/2015MNRAS.446.1252K} {446, 1252}

\bibitem[\protect\citeauthoryear{Rabiner \& Gold}{Rabiner \&
  Gold}{1975}]{Rabiner:1975p695}
Rabiner L.,  Gold B.,  1975, Theory and application of digital signal
  processing.
Prentice-Hall signal processing series, Prentice-Hall

\bibitem[\protect\citeauthoryear{Reichold et~al.}{Reichold
  et~al.}{2006}]{Reichold:2006p154}
Reichold A.,  et~al., 2006, {Particle accelerator. Proceedings, 10th European
  Conference, EPAC 2006, Edinburgh, UK, June 26-30, 2006}, C060626, 520

\bibitem[\protect\citeauthoryear{{Seiffert}, {Mennella}, {Burigana},
  {Mandolesi}, {Bersanelli}, {Meinhold}  \& {Lubin}}{{Seiffert}
  et~al.}{2002}]{2002A&A...391.1185S}
{Seiffert} M.,  {Mennella} A.,  {Burigana} C.,  {Mandolesi} N.,  {Bersanelli}
  M.,  {Meinhold} P.,   {Lubin} P.,  2002, \mn@doi [\aap]
  {10.1051/0004-6361:20020880}, \href
  {http://adsabs.harvard.edu/abs/2002A%26A...391.1185S} {391, 1185}

\bibitem[\protect\citeauthoryear{Stevenson}{Stevenson}{2013}]{stevenson_thesis}
Stevenson M.~A.,  2013, PhD thesis, California Institute of Technology,
  Pasadena, California, \url
  {http://resolver.caltech.edu/CaltechTHESIS:09292013-182521898}

\bibitem[\protect\citeauthoryear{{Trippe}}{{Trippe}}{2014}]{2014JKAS...47...15T}
{Trippe} S.,  2014, \mn@doi [Journal of Korean Astronomical Society]
  {10.5303/JKAS.2014.47.1.015}, \href
  {http://adsabs.harvard.edu/abs/2014JKAS...47...15T} {47, 15}

\makeatother
\end{thebibliography}
